\documentclass[usenatbib]{mn2e}

\def\msol{\hbox{$\rm\thinspace M_{\odot}$}}
\def\msolyr{\hbox{$\rm\thinspace M_{\odot}$ yr${}^{-1}\;$}}
\def\etal{{\it et al.\thinspace}}
\def\eg{{\it e.g.\ }}

\def\fig{Figure }

\def\p3m{P${}^3$M}
\def\ap3m{AP${}^3$M}
\def\-{{\em{---}}}

\usepackage{amsmath}       
\usepackage{amssymb}
\usepackage{wasysym}	   
\usepackage{mathrsfs}      
\usepackage{graphicx}      
\usepackage{setspace}       
\title[AGN Feedback models: Star formation and time evolution]{AGN 
Feedback models: Correlations with star formation and observational
implications of time evolution}
\author[Robert J. Thacker, C. MacMackin, J. Wurster and Alexander Hobbs]{Robert J. 
Thacker${}^{1}$\thanks{E-mail:thacker@ap.smu.ca}, C.
MacMackin${}^{1}$, James Wurster${}^{1,2}$ and Alexander Hobbs${}^{3}$\\
$^{1}$Department of Astronomy \& Physics, Saint Mary's University, Halifax, NS, B3L 3C3, Canada\\ $^{2}$Current 
address: Monash Centre 
for Astrophysics, Monash University, Victoria, 3800, Australia\\
$^{3}$Institute for Astronomy, ETH Zurich, Zurich, 8093, Switzerland}
\begin{document}

\date{Submitted March 5, 2014}

\pagerange{\pageref{firstpage}--\pageref{lastpage}} \pubyear{2013}

\maketitle

\label{firstpage}


\begin{abstract}

We examine the correlation between the star formation rate (SFR) and 
black hole accretion rate (BHAR) across a suite of different AGN 
feedback models, using the time evolution of a merger simulation. By 
considering three different stages of evolution, and a distinction 
between the nuclear and outer regions of star formation, we consider 63 
different cases. Despite many of the feedback models fitting the 
$M-\sigma$ relationship well, there are often distinct differences in 
the SFR-BHAR correlations, with close to linear trends only being 
present after the merger. Some of the models also show evolution in the 
SFR-BHAR parameter space that is at times directly across the long-term 
averaged SFR-BHAR correlation. This suggests that the observational 
SFR-BHAR correlation found for ensembles of galaxies is an approximate 
statistical trend, as suggested by Hickox et al. Decomposing the SFR 
into nuclear and outer components also highlights notable differences 
between models and there is only modest agreement with observational 
studies examining this in Seyfert galaxies. For the fraction of 
the black hole mass growth from the merger event relative to the final 
black hole mass, we find as much as a factor of three variation among 
models. This also translates into a similar variation in the 
post-starburst black hole mass growth. Overall, we find that while 
qualitative features are often similar amongst models, precise 
quantitative analysis shows there can be quite distinct differences.

\end{abstract}

\section{Introduction}
\label{intro}

A growing body of observational evidence (for a recent review see \citealp{AH12} and references 
therein) 
suggests that the growth of supermassive black holes (SMBH) is intrinsically linked to
properties of the host galaxy. Yet these relationships, be they correlations of bulge 
properties such as 
mass, luminosity or velocity dispersion relative to the black hole mass 
(e.g. \citealp{Metal98,Getal00,FM00,Gul09}) or the similarity of cosmological SFR and BHAR histories
(e.g. \citealp{BT98,Sil08,A10}), 
are 
subtle and not easily understood. While at a general level numerous mechanisms are known for
fueling black hole mass growth, such as galaxy mergers (e.g. \citealp{San88,SDH05,Hop06}), determining 
precise predictions for theoretical models remains a challenge because of the inherent difficulty in
understanding both accretion down to the SMBH scale (e.g. \citealp{Sh89,HQ10}) and the accompanying 
energy 
release 
ubiquitously known 
as AGN `feedback' (e.g. \citealp{SR98,K03,PK04,SDH05,O10}).

While, as noted, there appears to be a strong correlation between the 
cosmological histories of SFRs and BHARs, on an individual object basis 
the correlation is less clear. Some observations have found positive 
correlations between SFRs and BHARs (e.g. \citealp{Lu08,Ser10,Bon11}), while others
have found flat 
or negative correlations (\citealp{Page12,Har12}).
However, AGN have a much shorter variability time scale than global star 
formation (e.g. \citealp{HH09}), meaning that any anticipated 
correlations may only become clear when averages over populations, which
will capture the rapidly accreting objects, are 
considered. Results presented in \cite{Chen13} for star 
forming galaxies appear to provide support for this assertion. For 
simulation work it is possible to average over outputs taken at 
different times thereby averaging over different evolutionary phases.
  
As well as considering global star formation in galaxies, observations 
have also focused on whether correlations are stronger with nuclear 
(roughly the sub-kpc scale) or extended star formation (e.g. 
\citealp{K07,DSR12,Lam13}). As might be expected, nuclear star formation correlates more strongly with 
black hole accretion, while star formation in the outer regions of 
galaxies shows a weak correlation (at least for the sample of Seyfert 
galaxies considered in \citealp{DSR12}). For the highest luminosity 
systems, specifically QSOs, such a division remains beyond observational 
techniques.

Many models of AGN feedback implemented with galaxy formation simulations have been published 
(e.g. \citealp{SDH05,SS06,TSC06,ONB08,BS09,DQM11}). Simulation techniques have reached the point 
where there is comparatively little difference between the resolved scales of simulations 
but once sub-grid modelling is introduced there can be significant variations 
between 
models (e.g. \citealp{WT13p,WT13c}, hereafter WT13a, WT13b; \citealp{NK13,Bar13,Rag13,Cos13,Hay13,Tes14}).
Incorporating the full temporal and spatial scales relevant to AGN is 
clearly something beyond the current capability of simulations, despite new simulations reaching 
impressively high resolution (e.g. \citealp{GB13}). However, the 
working hypothesis for the field is that resolution down to the pc level is enough to capture most 
of 
the relevant physics (e.g. \citealp{HQ10}) and that augmenting sub-grid models to include 
different radiative behaviours of the SMBH may be the key step forward. New 
models 
which 
attempt to incorporate both so-called 
radio mode and quasar-mode feedback are now appearing \citep{V13}. 

To date, most simulation models have ignored the timescales associated
with accreting material on to the black hole. The
accretion-disc-particle model of \cite{PNK11} addresses this issue
partially, by considering the viscous time associated with the black
hole accretion disk. The original motivation of the model was to 
demonstrate the	importance of angular momentum to the accretion	process,
something that Bondi-Hoyle models do not account for. While this model was
originally intended for simulating accretion on pc scales, it has been
shown to produce acceptable results for merger simulations (WT13a), but
not cosmological \citep{Mul13}, and been further modified by \cite{NK13}
to incorporate the time-scale associated with material reaching the
accretion disc. Along similar lines a recent preprint (Rosas-Guevara et.
al. 2013, submitted) has attempted to incorporate the viscous timescale
associated with accretion by considering the circularization radius to
be determined by the flux of angular momentum through the smallest
resolved simulation scale.

Attention has also focused on more accurately describing the accretion 
processes within the wider galactic potential. \cite{HPNK12} have 
demonstrated that the accretion rate can be influenced strongly by the 
presence of additional mass beyond the black hole (from the stellar 
bulge, for example, or indeed the more massive dark matter halo) when 
the gravitational potential energy dominates over the internal energy of 
the infalling gas. However, as yet, there has not been a published study of 
this model, and hence we undertake one as part of this investigation and 
include it within our comparison.

Taking all these issues together, the aim of this investigation are to extend
the understanding of the SFR-BHAR correlation in context of different AGN
feedback models. The specific goals are:
\begin{itemize} 

\item Measure the intrinsic time variation of a single merger event, and 
thus quantify time variation in the SFR-BHAR parameter space. While not 
equivalent to ensemble averaging, it quantifies the evolutionary 
variation of a single AGN formation event (see section \ref{sfrbhar} for 
a discussion).

\item Calculate the SFR-BHAR correlations for this merger, considering 
evolution across all the simulation, and both pre- and post-merger 
cases. Contrast the different models, including those with explicit 
accretion timescales such as the \cite{PNK11} model, to observed 
correlations to see what can be inferred. 

\item Evaluate the SFR-BHAR 
correlations both for nuclear and extended star formation regions to see 
if observational expectations are matched. Because AGN accretion 
and nuclear star formation are both fed by cold gas in the nuclear 
region stronger correlations between nuclear star formation and the BHAR 
are expected. 

\item Extend our model framework to include the 
\cite{HPNK12} model in the merger context. 

\end{itemize}

The layout of this paper is as follows: In section 
\ref{sfrbhar} we discuss our handling of evolutionary tracks in 
the SFR-BHAR parameter space. In section \ref{sims} we review the 
numerical methodology and simulations, and follow this with a detailed 
in analysis in \ref{results}. In section \ref{conclusion} we conclude 
with a brief review.


\section{Galaxy evolution in the SFR-BHAR parameter space}\label{sfrbhar}

As galaxies evolve their instantaneous SFR and BHAR chart an 
evolutionary 
track in the SFR-BHAR parameter space. While observationally it is only 
possible to reconstruct these tracks in an averaged sense 
\citep{Wild10}, for simulations the evolution can be plotted exactly.

What can we learn from this analysis? The key issue is generating an 
underlying qualitative understanding of the SFR-BHAR correlation. If 
galaxies track along this correlation, both in an exact or time-averaged 
sense tightly, then the relationship is strongly suggestive of an 
evolutionary explanation. If, on the other hand, the evolutionary tracks 
run diametrically opposite the correlation then a large intrinsic 
scatter is to be expected. 


To put the simulation results in context it is instructive to examine 
what kind of behaviour in the SFR-BHAR parameter space might be expected. Let us first assume
an SFR based upon a Lagrangian Schmidt Law,
\begin{equation}
\dot{M}_* = {C_{sfr} \rho^{1/2}_g M_g },
\end{equation}
where $C_{sfr}$ is dimensional constant that can be related to the star formation
efficiency, 
$M_g$ and $M_*$ are the amounts of gas and stars in a given Lagrangian region and $\rho_g$ is
the gas density. For the
black hole accretion rate we utilize the Bondi-Hoyle accretion formula,
\begin{equation}
\dot{M}_{Bondi}  = {4 \pi G^2 M^2_{BH} \rho_{\infty} \over (c^2_{\infty} + v^2)^{3/2} }, 
\end{equation}
where $\rho_\infty$ and $c_\infty$ are the gas density and sound speed at infinity, 
$v$ is the relative velocity between the gas at infinity and the black hole and $M_{BH}$ is 
the mass of the black hole. Both systems are 
self-limiting in closed box situations. The mass in the stars, and 
equivalently the black hole 
mass, can only convert as much material as is available in a given gas 
reservoir. 

To determine what kind of behaviour is possible, first consider early
evolution in galaxies without a significant bulge, that are nonetheless gas
rich. We imagine a merger will occur and form an elliptical system at a later time.
Note, for simulations, the values at infinity are usually calculated as values
in the vicinity of the black hole sink particle.
For the outlined scenario, the $M_{BH}^2$ dependence means that the BHAR will be 
low and evolving slowly in an absolute sense, even though the relative change in mass in a given
time period,
i.e. $\Delta M_{BH} / M_{BH}$, can be significant. 
The SFR at this stage is as high as it can be
 and the trend is towards lower
SFR values with time. 
Hence at early stages, we expect to see small changes in the SFR and a comparatively constant, but slightly
rising BHAR producing a movement to the left and perhaps slightly upward in the SFR-BHAR space.

Jumping next to the final stages of evolution, as the gas is essentially 
exhausted, we can examine this under a closed box, fixed volume 
situation. It's also worth emphasizing that there is clearly a 
distinction between the nuclear gas supply and that available for star 
formation in the rest of galaxy. If the initial gas mass in the galaxy 
is $M_{g_{i}}$, then
the SFR behaviour at fixed volume is proportional to $(M_{g_{i}}-M_*)^{3/2}$,
which produces an inverse cube reduction in the SFR with time. For 
the BHAR, as the remaining gas mass becomes exhausted, but was initially 
given by $M^{nuc}_{g_{i}}$, then the BHAR is proportional to 
$M^{nuc}_{g_{i}}-M_{BH}$, which produces an exponential turn-off in time. 
Thus at late times, or whenever nuclear gas to feed the black hole
is exhausted, the slope in the SFR-BHAR 
parameter space can be expected to be steep due to the exponential
turn-off in the BHAR.

\begin{figure}

\begin{center} 
\includegraphics[width=1.0\columnwidth]{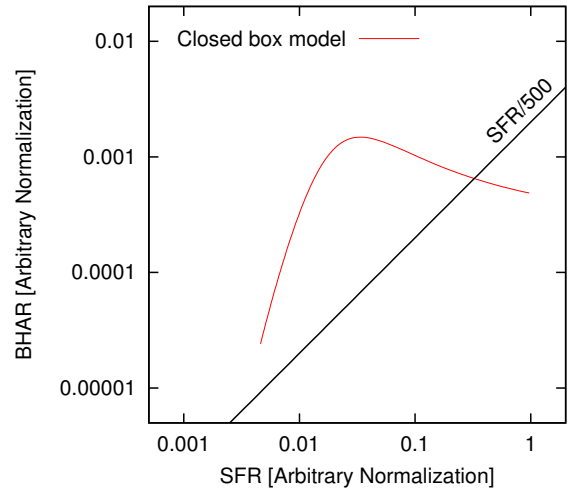} 
\end{center} 
\caption{Evolution of the SFR and 
BHAR for a closed box model of both the AGN and galaxy, plotted in the SFR-BHAR parameter space. Normalizations 
are arbitrary, as is the time at which the peak of the BHAR occurs, but the values have been chosen to 
approximately correspond to the values found in the merger simulations presented later. The SFR/500 line is 
included for reference with later plots, see section \ref{evo}. \label{tmodel}} 
\end{figure}

Thus the expectation is shallow evolution at the beginning, a rising BHAR,
followed by a steep turn-off. 
We have confirmed this behaviour by creating a toy-model of the 
closed box situation. Both the SFR and BHAR equations have analytic 
solutions, although the BHAR solution is implicit. We set the peak of 
the BHAR rate (which can be chosen by setting constants in the implicit 
solution) to occur halfway through the evolutionary period, slightly 
earlier but comparable to the simulations we present. We have also 
normalized the SFR to unity initially and chosen the mass associated 
with the nuclear region to be 0.01 times the mass of the galaxy. The 
resulting SFR-BHAR evolutionary track is shown in Figure \ref{tmodel}, 
and confirms the earlier presumptions. This model is, however, a 
significant oversimplification. It ignores mass flux between the galaxy 
and nuclear regions, the impact of self-gravity, changes in the sound speed and the impact of 
feedback. Nonetheless, the exponential turn-off result (in the absence of new
fuel for the BH) and the early evolution toward higher BHAR values appear
well motivated. 

It should be noted that a single merger simulation is obviously not 
enough to probe the SFR-BHAR correlation on an ensemble of galaxies. 
Current cosmological simulations (\eg \citealp{Ang13}) are already 
comparing to the \cite{Chen13} results, but future large-scale 
simulations with larger volumes (i.e. box sizes in excess of 40 
h${}^{-1}$ Mpc) will be able to do this analysis full justice. Once 
performed, they will allow a precise quantitative calculation of the 
scatter around the SFR-BHAR as well as potentially uncovering any 
expected changes in that correlation with galaxy type and age as well as 
cosmological redshift.

In constructing a track, the need for the inclusion of mergers means 
that decisions must be made on how to handle the precise definition of 
the pre-merger SFR and BHAR. We have chosen to take an average of the
identical galaxies prior to the merger but the differences to 
the other methods, such as choosing just one galaxy, or adding the SFRs and BHARs, amounts to only
a factor of two difference. Since the evolution moves through orders of magnitude changes
this is not a significant issue. 


Lastly, on a note of clarification, we use 
pre-merger to describe the systems up until 980 Myr of evolution, just 
at the epoch of the core merger but before a large amount of material 
flows into the nucleus. This means that our definition of post-merger 
necessarily includes data for the very highest SFRs and BHARs and might 
more appropriately be considered ``merger and post-merger".

\section{Numerical models}
\label{sims}
\subsection{Merger and galaxy model}
\label{simsGM}

\begin{table}
\caption{Particle and galaxy component masses for the modelled galaxies.\label{comps} }
\begin{tabular}{l|r r r r}
\hline
 & Total mass                 & Particle mass           & Number of  \\
Component & (10$^{10}$M$_{\astrosun}$) & (10$^5$M$_{\astrosun}$) & particles  \\
\hline
Dark matter halo & 89.92    & 11.75 & 765 000  \\
Hot gas halo     &  0.60    &  0.36 & 165 343  \\
Stellar bulge    &  1.34    &  2.37 &  56 649  \\
Stellar disc     &  3.56    &  2.37 & 150 375  \\
Gas disc         &  0.54    &  0.36 & 150 375  \\
\hline
\end{tabular}
\end{table}

Full details of the galaxy models, which are Milky Way analogues, may be 
found in WT13b. In Table \ref{comps} we give component breakdowns of the 
halo, bulge and disc components of the galaxies at the fiducial 
resolution we consider here.

Our approach to modelling radiative cooling is
the same as \cite{WT12}, which includes a representation of cooling down to 300 K, using the tables
of \cite{WN01}. In all simulations, radiative cooling is implemented using an assumed 
metallicity of $Z = 0.05Z_{\astrosun}$ which means that increasing metallicity due to enrichment
is not explicitly tracked. This simplifying step, which admittedly does omit some notable physics, allows us to avoid logarithmic changes in the
cooling curves in small areas, which are extremely difficult to track accurately (see \citet{TC92} for a discussion
of the challenges of modelling cooling accurately in SPH calculations with strong density gradients).

\subsection{Star formation}

As noted in WT13b, the star formation algorithm used in all these 
simulations \citep{tc00} is kept constant to minimize differences from one 
AGN feedback algorithm to another. The algorithm is based upon the 
``classical" approach that enforces the star formation rate by utilizing 
a Lagrangian version of the Schmidt Law (e.g. \citealp{K92,K98}). The 
method also relies upon the assumption of pressure equilibrium between 
ISM phases to estimate the density of local gas that should be 
transferred to a hotter phase under feedback. This approximate density, 
which is much lower than that of cold gas, is then used in the cooling 
function because the SPH density responds more slowly than the cooling 
times of high density gas ($n_H> 1$ cm${}^{-3}$) at the typical feedback 
temperature employed ($10^6$ K). In practice this approach is very 
similar to delayed-cooling models (e.g. \citealp{SSKWGQ06}), which 
remain useful when simulations do no have enough resolution in the 
spatial or time domains to resolve energy input sufficiently accurately.

While this method is well documented in the literature, it has some 
notable differences to more recent approaches that rely upon effective 
equations of state (e.g. \citealp{SH03}). Perhaps the most well known difference is 
that the classical model has a stronger resolution dependence than EOS 
based approaches, although this must be tempered with recognition that 
increased resolution should allow for better capturing of local density 
gradients and hence the local SFR.

There are subtle differences as well. In particular, Springel et al. 
(2005) demonstrate that at the apoapsis of the merger there is rapid 
star formation in the classical approach, but not in the EOS-based 
version. This is due to the gas being kept dynamically hotter in the 
latter model, at both low and high resolution. It is also worth noting 
that the precise bar behaviour in these merger models is very sensitive 
to nuclear (bulge) masses. For example, we showed in WT13b that the 
inclusion of black hole tracer particles of mass $10^{9}$ \msol (as in 
the DQM model) was enough to stabilize against bar formation, while 
lighter tracer particles did not. This behaviour highlights the 
difficulty in correctly evolving instabilities in the presence of a 
dynamically changing potential.

Nonetheless, despite these documented differences, the SFR results we 
presented in WT13b are in broad agreement with the other works that used 
EOS-based approaches and the differences between AGN models produced 
notably larger impacts on the SFR values.

\subsection{AGN Feedback implementations}

We revisit the numerical models first discussed in WT13a, WT13b and also add a new model 
implementing the Hobbs et al (2012) algorithm. In our previous paper we highlighted that there
are essentially five key attributes to an AGN feedback implementation:
\begin{enumerate}
  \item The accretion rate on to the black hole,
  \item The SPH particle accretion algorithm,
  \item The energy feedback algorithm,
  \item The black hole advection algorithm, and
  \item The black hole merger algorithm.
\end{enumerate}
For completeness (additional details may be found
in WT13b) we summarize salient features of each algorithm, in the context of the above
attributes. It is worth emphasizing that there is a strong distinction between the numerical 
influence radius of the black hole, $r_{inf}$, often set to determine a fixed number of neighbour particles, as 
compared to the gravitational sphere of influence $r_h=GM_{BH}/\sigma^2$, where $\sigma$ is the 
local velocity dispersion.

\subsubsection{Model 1: SDH}

This model is based upon the model found in \cite{SDH05} (herein SDH05). The accretion
rate is given by a modified Bondi accretion rate,

\begin{equation}
\label{mBondi}
\dot{M}_{Bondi} = \frac{4\pi\alpha G^2 M^2_\text{BH} \rho}{\left(c_s^2 + v^2\right)^{3/2}},
\end{equation}
where $c_s$ and $\rho$ are the local sound speed and local density of the gas, and $v$ is the
relative velocity of the black hole to the nearby gas. A free parameter $\alpha$, which
we set to 100, is included to
adjust for the limited maximum density resolved in these merger simulations.
The maximum accretion rate is limited by the Eddington rate,
$\dot{M}_\text{BH} = \min\left( \dot{M}_{Bondi}, \dot{M}_{Edd} \right)$. To accrete particles on to the black
hole a ``stochastic-unconditional"
algorithm is used. Particles within $r_{inf}$ of the black hole particle are tested against a calculated
probability of accretion based upon the black hole growth rate and the local density. Note that while $M_{BH}$
denotes the mass of the black hole in the model (frequently referred to as the `internal' mass), the actual dynamical mass in the simulation, $m_\text{BH}$, builds up over
time by particle accretion and can be slightly different from $M_\text{BH}$. Of course, an accretion algorithm
should ideally maintain $m_\text{BH}\sim M_\text{BH}$ (see WT13b). 

Energy is returned to particles within $r_{inf}$ using a coupling efficiency of 5\% and assuming
an overall energy output of $\epsilon_r \dot{M}_\text{BH} c^2$ with the radiative efficiency $\epsilon$ 
set to 10\%. Energy is returned isotropically and is also weighted by the local
SPH kernel so that particles further from the black hole particle receive less energy.

Advection of the black hole proceeds by moving the position to the gas particle with the lowest
potential provided that the relative velocity between them is less than one quarter of the local 
sound speed. Black hole mergers occur when two black hole particles come within their mutual
SPH smoothing lengths and their relative velocity is less than the local sound speed.
 
\subsubsection{Model 2: BS}
Primarily designed for cosmological volumes, this model (see \cite{BS09}, hereafter BS09) builds upon   
the SDH implementation by modifying the $\alpha$ parameter to produce higher feedback when
the local density goes above a threshold density, $n_H^*$, of 0.1 cm${}^{-1}$. The $\alpha$ parameter thus becomes
a function of the local hydrogen density,
\begin{equation}
\label{alpha}
\alpha = \left\{ \begin{array}{c l} 1 		    & \text{if   } n_\text{H} < n_\text{H}^* \\
\left(\frac{n_\text{H}}{n_\text{H}^*}\right)^\beta  & \text{otherwise}
\end{array}\right., 
\end{equation}
and following BS09 we have set $\beta=1$. As with SDH the maximum accretion rate is also Eddington limited.

The feedback energy in this model is calculated in the same way as SDH, but the 
coupling efficiency is taken to be three times higher ($\epsilon_f$=0.15, 
$\epsilon_r=0.1$). The energy is returned to particles individually though, rather than spread
over neighbours once a critical energy is reached, given by,

\begin{equation} \label{Ecrit} E_\text{crit} = \frac{m_\text{g} k_\text{B} \Delta 
T}{\left(\gamma - 1\right) \mu m_\text{H}}, 
\end{equation} 
where $m_\text{g}$ is the (initial) 
mass of a gas particle and $\Delta T$ is the temperature increase a particle experiences with 
every feedback event. We set a lower temperature threshold of $5\times10^6$ K to primarily ensure
stability of integrations at our mass resolution, which is appreciably higher than that used in BS09.
This choice leads to more frequent but less powerful episodes of feedback, meaning that the amount
of hot gas in the halo may possibly be lower in our models. However, over the lifetime of the merger
the feedback energy budgets should be similar regardless of the chosen $\Delta T$.

Gas particles are accreted by a stochastic-conditional particle accretion algorithm.  If 
$M_\text{BH} < m_\text{BH}$, then the probability of accretion is $p_i \equiv 0$, otherwise it 
is calculated using the mass difference, the local density and the kernel weight $w_i$, via, 
\begin{equation}
\label{prob2}
p_i = w_i \left(M_\text{BH} - m_\text{BH}\right)\rho^{-1}.
\end{equation}
As in Model SDH, particle $i$ is accreted if $p_i > x_i$, where $x_i$ is a random number.

The black hole advection is the same as in Model SDH, while two black holes are considered
to have merged when they 
come within each other's smoothing lengths and have a relative velocity less than the circular 
velocity at the radius of the most massive black hole's smoothing length. 

\subsubsection{Model 3: ONB}

This model was also originally developed \citep[hereafter ONB08]{ONB08} for use in simulations using cosmological 
initial conditions. It is also solely focused on reproducing radio-mode feedback rather than 
the brighter quasar-mode so different behaviours should be expected for it, and indeed are 
found (WT13b). Based upon the model of \cite{KU02} mass growth of the 
black 
hole is determined by radiative drag estimates on the ISM near the black hole leading to 
a loss of angular momentum 
and accretion. The net accretion rate from this drag is given by,
\begin{equation}
\label{drag}
\dot{M}_\text{drag} = \epsilon_\text{drag} \frac{L_\text{RSF}}{c^2}\left(1-\text{e}^{-\tau_\text{RSF}}\right),
\end{equation}
where $\epsilon_\text{drag} = 1$ is the drag efficiency, $L_\text{RSF}$ is the total bolometric 
luminosity of all the stars in the region of star formation (RSF) near the black hole, and $\tau_\text{RSF}$ is 
the total optical depth of the RSF. Luminosities are found using {\sc pegase2} \citep{FR97} while the optical
depth is calculated from the total mass of clouds in the RSF, its radius and the mass extinction coefficient.

Feedback in this model is directed specifically into the halo via a jet modelling approach, which also 
includes a distinction between jets from standard (optically thick) thin disks versus those from 
radiatively inefficient accretion flows, with optically thin but geometrically thick disks. Thermal 
energy associated
with the jets is distributed to the nearest 40 gas particles below a specified density threshold.

Particles are accreted via a probabilistic approach whenever the internal mass exceeds its dynamical 
mass, a process we call ``continual-conditional" accretion. The black hole trajectory always heads
toward the steepest stellar density via,
\begin{equation}
\label{deltal4}
\Delta l_\text{ONB} = \min(0.01\epsilon_\text{S2}, 0.03 \left| {\bmath{v}} \right| \mathrm{d}t),  
\end{equation}
where $\epsilon_\text{S2}$ is the gravitational softening length, ${\bmath{v}}$ is the 
velocity of the black hole, and $\mathrm{d}t$ is the time-step; these coefficients are the same as in 
ONB08 and were determined empirically. Mergers of black holes occur when both black hole particles are 
within their mutual softening lengths, and are gravitationally bound.

\subsubsection{Model 4: DQM}
This model \citep[hereafter DQM11]{DQM11} uses a fundamentally different approach to accretion, and focuses on the transport
of material from large scales to small via the ``instabilities within instabilities" concept and
gravitational torques \citep[e.g.][]{HQ10}.  The accretion rate is 
\begin{equation}
\label{Visc}
\dot{M}_\text{visc} = 3 \pi \delta \Sigma \frac{c_\text{s}^2}{\Omega}, 
\end{equation}
where $\delta$ is the dimensionless viscosity, $\Sigma$ is the mean gas surface density, and $\Omega = 
\sqrt{GM/r_\text{inf}^3}$ is the rotational angular velocity of the gas. We set the free parameter 
$\delta$ to 0.05 as in DQM11. 
 
Feedback energy is returned via a momentum approach, assuming an infrared optical depth of 10. The 
momentum is injected radially and isotropically on to the particles within $r_{inf}$. As in other models, 
the luminosity is limited by the Eddington rate such that $L = \min \left( \epsilon_\text{r} 
\dot{M}_\text{visc} c^2 , L_\text{Edd} \right)$. 

Black holes are modeled using tracer particles of mass $10^9$ \msol. This necessarily decouples the 
internal mass from any concept of a dynamical mass (as it is held fixed). The large masses of the black 
hole particles means they preferentially follow the local minimum of the potential. We randomly remove 
gas particles from the simulation that are within two smoothing lengths of the black hole to match the 
increase of the internal black hole mass. Mergers of black holes occur when they approach within one softening length of one another
regardless of their velocity.

\subsubsection{Model 5: WT}

This model (see WT13b) combines a number of approaches that have appeared in the literature to draw
together algorithms showing desirable behaviours (such as stability of the black hole trajectory). The 
modified Bondi accretion rate of SDH05 is used, but feedback energy is returned thermally using a top-hat 
kernel for all particles within $r_{inf}$. This prevents excessive heating close to the black hole.

Particle accretion is handled using a continual-conditional algorithm:  When $M_\text{BH} > 
m_\text{BH} + m_\text{g}/2$, we accrete the gas particle that is nearest to the black hole. This keeps
the internal and dynamical masses very closely coupled.

Black hole advection is broadly similar to that of ONB, but utilizes the total local potential rather than
just stellar particles. The distance the black hole is displaced has been modified to
\begin{equation}
\label{deltal5}
\Delta l_\text{WT} = \min(0.10h_\text{BH}, 0.30\left| {\bmath{v}} \right| \mathrm{d}t).
\end{equation}
Even in the presence of voids produced by winds this approach produces a smooth track for the black
hole trajectory. Mergers of black holes rely upon the same approach as SDH05. 

\subsubsection{Model 6: PNK}

The PNK model \citep{PNK11} couples together the black hole and associated accretion disk processes.
The original motivation behind the model was to address the issue that Bondi-Hoyle approaches
overestimate accretion of rotationally supported cold gas disks. The model also
was extended to include the viscous timescale, $t_{visc}$, associated with material accreting from the accretion disk
into the black hole. 

Particles are accreted onto the accretion disk, of mass $M_\text{disc}$, whenever they fall within the 
accretion radius $R_\text{acc}$, which is typically of order a few pc. It is worth noting
that this is of course far below the resolution scale of these simulations, so in practice $R_\text{acc}$
behaves as an accretion rate limiter. The mass that is extracted is then added to the 
accretion disk mass, which in turn accretes on to the black hole at a rate 
$\dot{M}_\text{BH}=\rm{min}(M_\text{disc}/t_\text{visc},\dot{M}_\text{Edd})$. From this accretion rate the feedback energy is 
returned using the same wind method as DQM. Black hole advection and mergers are handled in the same way as the WT model.

In WT13a we investigated a number of different accretion radii and viscous timescales. To make our current analysis
compact, we have decided to focus on a model that uses a 5 Myr accretion timescale and an accretion radius that
is 5\% of the minimum smoothing length (which essentially sets the mass flow rate onto the accretion disk).
We refer to this model throughout the paper as PNK0505, although in WT13a it is labelled as PNKr05t05. 

\subsubsection{Model 7: HPNK}

Even in the absence of significant angular momentum in accretion the 
Bondi-Hoyle formalism can still lead to inaccurate accretion rates, for 
example when gas can free-fall due to highly efficient cooling 
processes. Similarly the potential is assumed to be derived solely from 
the mass of the black hole, whereas in galaxies the surrounding halo 
could legitimately be expected to have an impact on the flow. 
\cite{HPNK12}, hereafter HPNK12, have shown that a modification of the 
Bondi-Hoyle formalism to include the enclosed total mass within the 
smoothing radius and the impact of the associated velocity 
dispersion, 
produces the interpolating formula

\begin{equation}
\dot{M}_\text{interp} = {4\pi \lambda(\Gamma) G^2 M^2_\text{enc} \rho_{\infty} \over (c_\infty^2 + \sigma^2)^{3/2}},
\end{equation}
which captures much of the desired behaviour. On small evaluation scales it 
approaches the Bondi-Hoyle formula while on larger ones it naturally 
includes the larger potential.

The smoothing radius in this model is set at the softening length
of 120 pc. We 
also keep this particular value fixed with time, rather than relying 
upon a variable value to enclose a certain number of neighbours as that 
could potentially change by large amounts if a cavity is blown in the 
gas distribution during violent feedback events. As discussed in 
\cite{Bar13} all black hole modelling approaches that use a variable 
smoothing length have the potential to develop voids, and we return to 
this point in section \ref{sec:nuc}.

For the feedback, black hole advection and merging algorithms we utilize 
those implemented in the WT model. This provides a direct
means of assessing the impact of changing the mass accretion rate.

\begin{figure}

\begin{center}
\includegraphics[width=1.0\columnwidth]{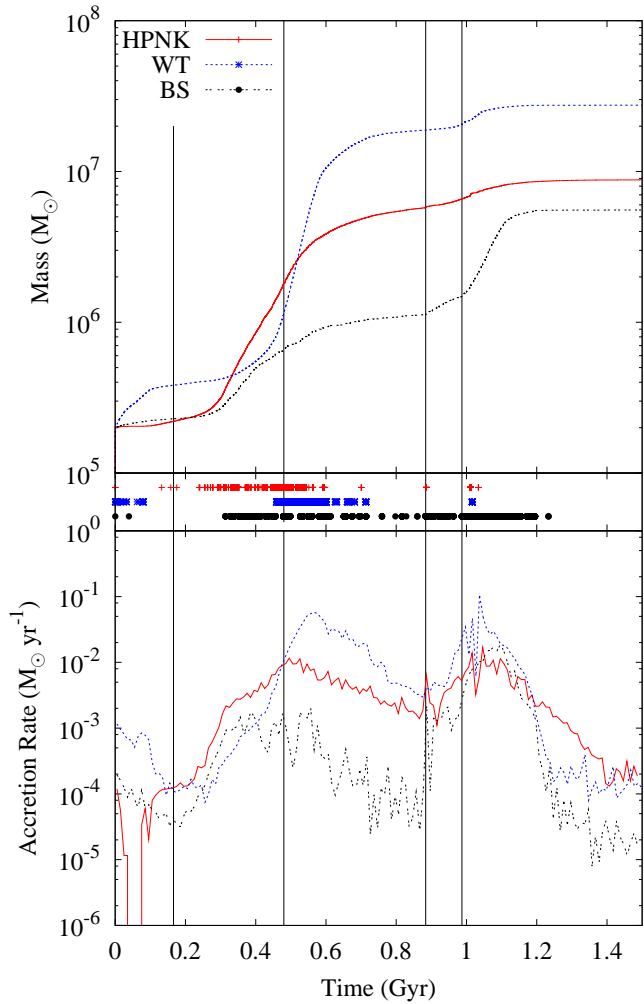}
\end{center}
\caption{Total black hole mass in the simulation versus time (10 Myr bins) for the BS, WT
and HPNK models (upper panel), epochs at which accretion is Eddington limited (middle panel), and the total accretion rate
on to the black holes (bottom panel). The black lines indicate, from left to right, first periapsis
at 166 Myr, apoapsis at 480 Myr, second periapsis at 884 Myr and the core merger
at 987 Myr.\label{HPNKmassev}}
\end{figure}

\section{Results}
\label{results}

With the exception of model ONB, all the models were evolved through the
merger to approximately 500 Myr afterwards, for a total simulation time
of 1.5 Gyr. The ONB model was only evolved for 1.25 Gyr due to a clustering
slowdown caused by the lack of feedback in this model. We still, however,
provide post-merger numbers for this simulation on the basis of the smaller
amount of data that we have. While the models have gravitational softenings
of 120 pc, and capture variations in star formation and AGN feedback over several 
orders of magnitude in density, it is unlikely that they are converged
with respect to small scale variations in the SFR and AGN feedback. We have, however,
shown that gross features, such as the black hole masses do seem to be predicted
well as a function of resolution in some models. Our primary focus is thus on
differences between models and qualitative trends, and we caution against
over-interpretation of the observational comparison.  

\subsection{HPNK compared to other models}

We first examine the impact on the black hole mass evolution which is given in the top panel of Figure 
\ref{HPNKmassev}. We compare to two other models, WT and BS, as the final mass of the 
WT model lies close to the \cite{Gul09} $M-\sigma$ relationship,
while the BS model has the lowest mass associated with Bondi-Hoyle type accretion models. Together these
two models give a good idea of the range of masses found in WT13b.
What is immediately noticeable is that the HPNK mass accretion rate has lead to 
significant mass growth prior to apoapsis. Specifically, at apoapsis the total mass in black holes for HPNK, WT 
and BS are respectively $1.80\times10^6$ \msol, $1.15\times10^6$ \msol and $0.67\times10^6$ \msol, making the 
black hole mass total almost 60\% higher in HPNK than WT. Examining the mass accretion rates, shown in the lower 
panel of Figure \ref{HPNKmassev}, shows that prior to apoapsis HPNK is almost always more active than the other 
two models plotted, although they do not have a specific ordering between themselves, with WT sometimes having 
higher rates than BS and vice versa. The higher accretion rates associated with HPNK also mean that there is 
more feedback occurring prior to and at apoapsis.

In the period between apoapsis and second periapsis, trends are notably different. While the accretion rate for 
WT rises sharply up to a peak at $\sim10^{-1}$ \msolyr and then falls, HPNK falls on a consistent trend to just 
above $10^{-3}$\msolyr and BS, as a result of strong feedback, is over an order of magnitude lower at second 
periapsis.

As the simulated galaxies begin to reach core-merger there is an increase in accretion for all the models. As a relative fraction of mass, between core-merger and the end of the simulation the total BS black hole mass grows by a factor of 2.55, while for WT the factor is 1.36, and for HPNK 1.20 (although see section \ref{merge} for a more detailed discussion of all
models). The final black hole mass for HPNK is $8.80\times10^6$ \msol and the stellar component has an associated velocity
dispersion of $144$ km s${}^{-1}$. For this $\sigma$ the mean of the \cite{Gul09} $M-\sigma$ predicts a final
black hole mass of $3.30\times10^7$ \msol, and, for their quoted standard deviation, the HPNK black hole mass is approximately 1.4 standard deviations below the mean. For comparison, the BS model, which has an essentially identical velocity dispersion, is over 2.1 standard deviations below the mean, while the WT model lies essentially on the mean.  

The SFR for HPNK is broadly similar to the BS model. There is only a moderate increase at both apoapsis and during the core merger, and the final stellar masses vary by only 2\%, with WT, BS and HPNK all being very close to $10.3\times10^{10}$ \msol. As might be expected given the similarity in the evolution, the final morphology is also similar to the BS model with the embedded central gas disk again being very small compared to other models such as WT. The hot circumgalactic gas halo is also smaller (when measured by a visual cuts at $10^6$ K and $10^5$ K) than WT or DQM, a result that is again similar to the BS model. 

\subsection{Ensemble correlations compared to time averaged simulations 
and SFR-BHAR evolution}\label{evo}

We first consider correlations of the SFR and BHAR for the entire 
galaxies by examining the evolutionary tracks and then taking time 
averages. In the absence of feedback, the closed box models discussed in 
section \ref{sfrbhar}
suggest that correlations in the SFR-BHAR space can become quite steep
after the peak BHAR is reached. However, prior to that epoch the
converse may be true, and shallow negative correlations are possible
depending upon the evolutionary epoch considered. As a general rule
feedback can be anticipated to shift both SFRs and BHARs to lower
values, but the precise impact on the exact correlation requires
evaluation from simulations.

In this analysis we follow the \cite{Chen13} convention of 
correlating the SFR against the BHAR, i.e. 
$\log{(BHAR)}=\alpha+\beta\times\log{(SFR)}$. We also further analyse 
the SFR behaviour by separating it into the nuclear and extended 
components, and to match the prior literature analyses, switch to 
correlating the BHAR against the SFR.

\begin{figure*}
\begin{center}
\includegraphics[width=2.0\columnwidth]{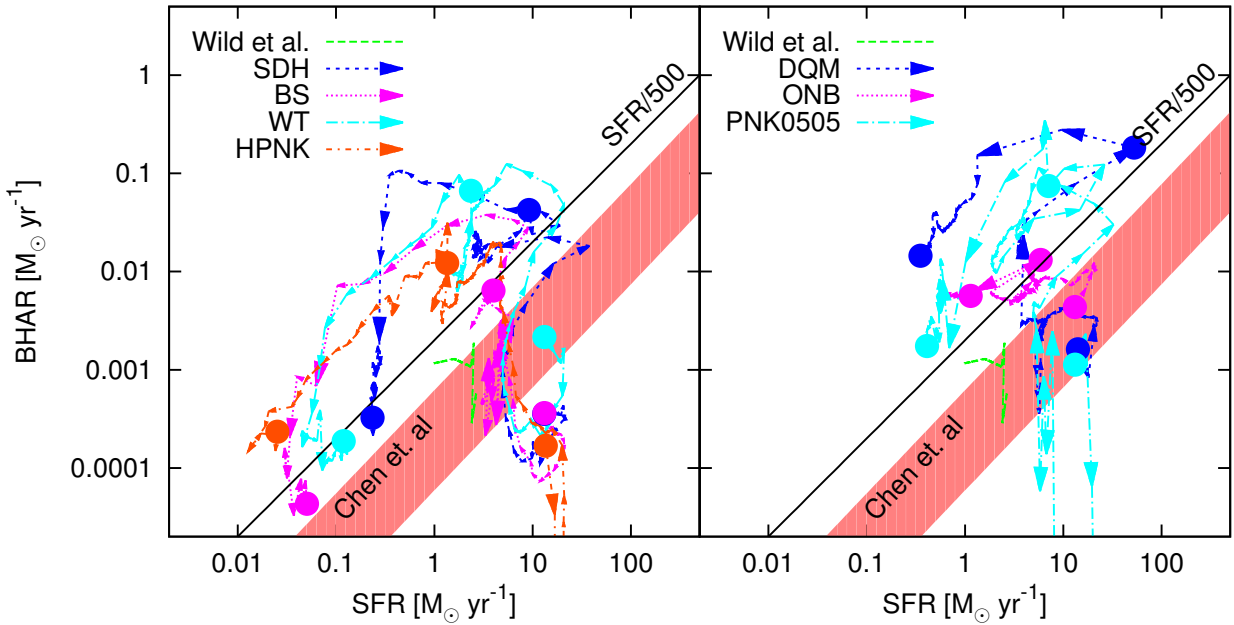}
\end{center}
\caption{Evolution of the different models in the BHAR-SFR parameter space. The left-hand panel displays models that rely on Bondi-Hoyle accretion or a variant of it, while the right-hand panel displays models using alternative accretion approaches. Also shown is the evolution inferred from
the Wild et. al. (2010) data, the Chen et al. (2013) correlation and the SFR/500 line derived from the $M_{BH}/M_{bulge}$ ratio found in Marconi et al. (2004). The evolution
steps are averaged over 20 Myr periods, with each arrow corresponding to evolution in that period. Longer arrows thus represent correspondingly larger movements in the parameter space for a given period of time. The beginning, core-merger at 980 Myr and final output at 1.5 Gyr are all marked with the appropriately coloured circle (note SDH and BS have overlapping
starting positions).
Most models evolve through a small loop down and then upwards towards the top right firstly as the initial conditions settle and later as the system reaches core-merger. This is followed by a trend to the lower left once the system has become starved of fuel for further star formation or black hole accretion.  
The notable exception to this behaviour is the ONB model which has a very narrow range in BHAR throughout the simulation.    
\label{sfr-bhar}}
\end{figure*}

\subsubsection{Evolution tracks in the SFR-BHAR parameter space}\label{evot}

By examining evolution in the SFR-BHAR parameter space we can we can 
gauge the overall variation between models and how this compares to 
observed properties such as inferred evolution \citep{Wild10}, measured 
correlations of infrared selected star-forming galaxies \citep{Chen13}, 
and the SFR/500 value derived from the $M_{BH}/M_{bulge}$ ratio found in 
\cite{Mar04}.

In \fig \ref{sfr-bhar} we plot the evolutionary tracks of the 
simulations in the SFR-BHAR parameter space along with the observational 
relationships. All the different models show similar variations in SFRs, 
over approximately two orders of magnitude (from 10s \msol\,yr$^{-1}$ to 
0.1 \msol\,yr${}^{-1}$), while the BHAR rates typically vary by around 
three orders of magnitude (from 0.1 \msol\,yr$^{-1}$ to $10^{-4}$ 
\msol\,yr${}^{-1}$). It is notable that the PNK models have a much 
larger variation in BHAR due to the exponential decay possible from the 
accretion disc reservoir.

In the log-log parameter space plot the evolution of the different 
models is qualitatively similar (with the ONB radio-mode behaving in
an expectedly different manner). The simulations start with a 
comparatively high SFR (10\msol\,yr$^{-1}$) due to the onset of cooling 
at the beginning of the simulation, and are accompanied by modest BHAR 
values ($10^{-3}$\msol\,yr$^{-1}$), placing the models in the lower 
right to middle of the parameter space. For some of the models (notably, 
SDH, BS, WT and HPNK) there is a small initial fall in the BHAR prior to 
first periapsis, but this is then followed by an increasing BHAR 
(slightly under 0.01\msol\,yr$^{-1}$) as the instabilities promoted 
during the initial pass are excited. This produces evolution that loops 
upwards towards the upper right of the diagram. 

During the core merger 
both the BHAR and SFR rise although the increase in 
SFR can be quite weak depending upon the strength of the AGN feedback and
the amount of gas available for star formation. At core-merger 
BHAR values range between 0.01\msol\,yr$^{-1}$ and 0.1\msol\,yr$^{-1}$, while
the SFR values range between 1\msol\,yr$^{-1}$ and 10s \msol\,yr$^{-1}$. Post merger, the 
systems decline in both the SFR and BHAR value and there is a trend 
diagonally down and left, which follows a close to linear correlation 
for a number of models (e.g. BS, WT, DQM and especially HPNK), albeit at a higher
normalization than either the \cite{Chen13} band or SFR/500 line.

In the ONB model the 
AGN feedback energy is channelled directly into the halo, and does not impact
the nuclear gas or SFR significantly (the SFR varies by approximately 1.5 orders
of magnitude). This lack of nuclear feedback, combined with the assumed accretion model,
means that the BHAR is comparatively
constant over time. While this behaviour is very different from other models,
it is interesting to note that ONB spends more time 
closer to the average track predicted by the \cite{Wild10} data than any 
of the other simulations although it is difficult to draw quantitative 
conclusions from this similarity.

\subsubsection{Time-averaged correlations for the entire galaxies}

We next quantify the evolution by evaluating correlations between the 
SFR and BHAR and compare directly to the ensembled-derived \cite{Chen13} 
relationship with $\beta=1.05\pm0.33$. Time averages weight all points along the evolutionary track 
equally and are less impacted by sudden rapid movements in the parameter 
space. Given the complexity of the evolutionary tracks over the entire 
merger simulation, it is clear that linear relationships, as in 
\cite{Chen13}, are unlikely to be recovered. However, breaking the 
evolution into pre- and post-merger provides a helpful subdivision as it 
isolates similar evolutionary epochs. In the analysis below we consider 
average values and ranges across the models, although we don't suggest 
that the average value across models has any specific meaning, rather it 
identifies trends in the correlations across the different models at 
different epochs.

\begin{figure*}
\begin{center}
\includegraphics[width=2.\columnwidth]{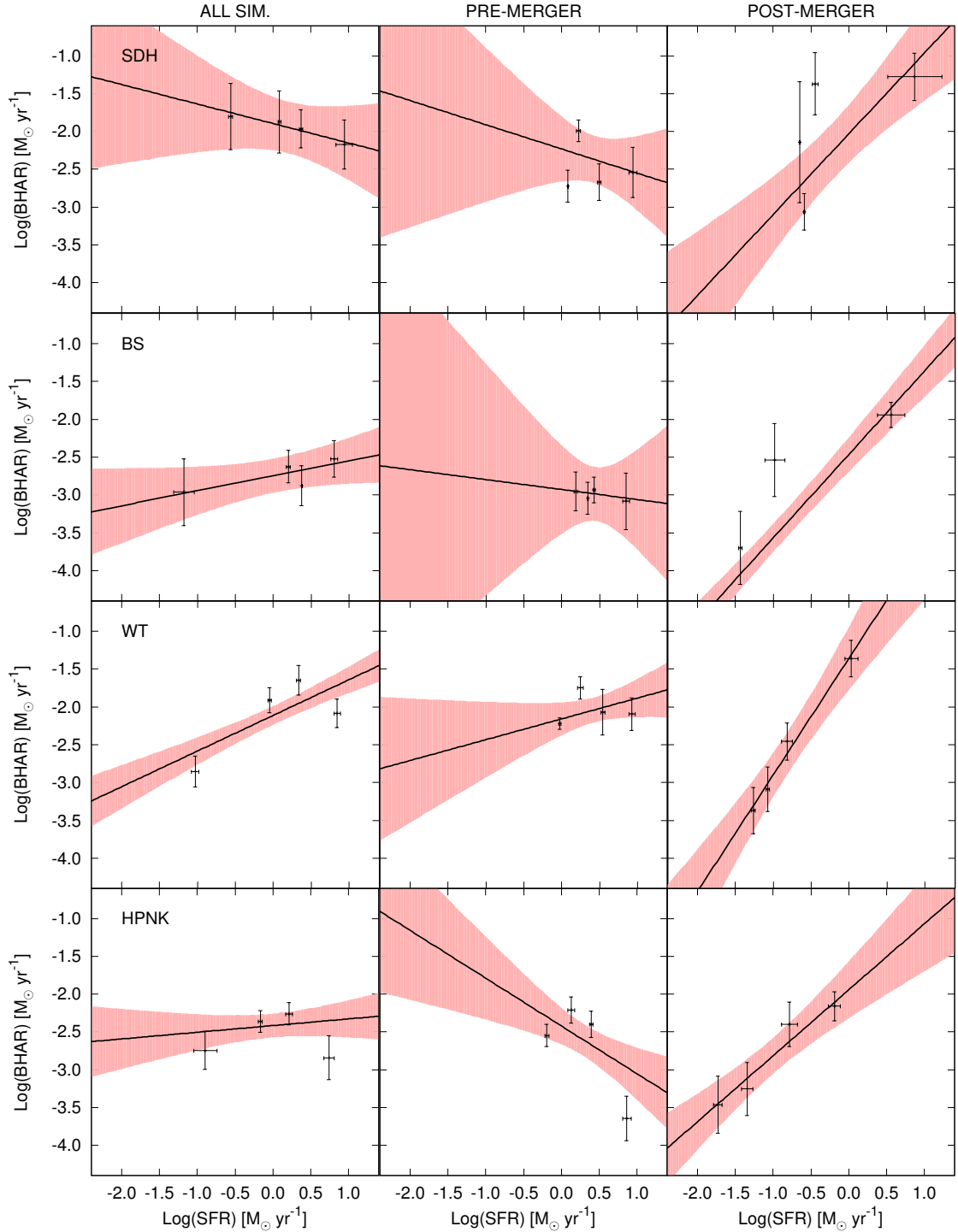}
\end{center}
\caption{Time averaged correlations between the total SFR and BHAR for the models that use Bondi-Hoyle
accretion or some variant of it. Results are given for three different epochs, all the simulation, 
the pre-merger evolution and the post-merger evolution. 
Error bars correspond to variances in the means and the dotted lines denote 95\% confidence bands around each fit.
Four points are given in each plot and correspond to match the binning approach chosen in Chen et. al. (2013),
with the values used in each bin coming from simulation outputs spaced 5 Myr apart (10 Myr in the case of HPNK). \label{bins1}}    
\end{figure*}

\begin{figure*}
\begin{center}
\includegraphics[width=2.\columnwidth]{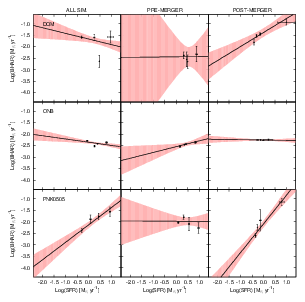}
\end{center}
\caption{Time averaged correlations between the total SFR and BHAR for the models that do not use the Bondi-Hoyle
approach to calculating accretion. All other details are as in Figure \ref{bins1}. \label{bins2}}
\end{figure*}

We use outputs exactly spaced 5 Myr apart to calculate correlations, 
except HPNK which, due to disc space limits were 10 Myr apart. As in 
 \cite{Chen13} we construct four bins in the SFR, however, we use a constant sample size 
in each bin, and the bin means are calculated via an arithmetic mean (see their 
equation 4). Variances in the means are then bootstrapped and used in the 
$\chi^2$-minimization linear (log space) regression fitting. The fits that we find, 
along with their standard errors and $\chi^2$ values, are summarized in Table 
\ref{all} and plotted in Figures \ref{bins1} and \ref{bins2}.
 
Examining the fits across the entire 
simulation given in Table \ref{all} we find that some $\chi^2$ values 
are poor (particularly DQM, ONB and PNK0505). Visual examination shows
that residuals can be tub shaped in many cases, which is to be expected given
the shape of the evolutionary tracks. In terms of the fitted 
slopes, the mean and range across all simulations is given by 
$\beta=0.12^{+0.64}_{-0.38}$ although most models fall around zero, with 
SDH, DQM and ONB being slightly negative, while BS, WT, HPNK and PNK0505 
are positive.

For the pre-merger evolution, there is a smaller spread in fitted 
slopes, with the mean and range (ONB neglected as having little 
variation) being $\beta=-0.08^{+0.36}_{-0.55}$. This indicates that 
rapid evolution up to higher BHAR values just prior to the core merger 
does not influence the time average significantly.  The WT model is the 
notable positive slope outlier ($\beta=0.28\pm0.22$) while the HPNK 
model has the largest negative slope at $\beta=-0.63\pm0.50$ although 
the standard error is large. The HPNK fit is clearly influenced by the 
right-most point that comes from the very low accretion values that are 
possible in this model. 
  
For the post-merger evolution, with the exception of ONB, all the models 
have positive slopes, with a mean and range of 
$\beta=0.91^{+0.63}_{-0.92}$. If we neglect the ONB model as a 
significant outlier, then the mean and range is notably tighter at 
$\beta=1.07^{+0.47}_{-0.46}$. Most of the models using Bondi-Hoyle 
accretion approaches (SDH, BS, WT, HPNK) all produce slopes close to, or 
above, unity, as does PNK0505. DQM has a somewhat less steep slope at 
$\beta=0.61\pm0.24$, but this value is still considerably steeper than 
either the pre-merger or full evolution values for this model. Thus with 
the exception of ONB, and within the bounds of error, all the 
post-merger correlations match the \cite{Chen13} power law.

\begin{table}
\caption{The time averaged total SFR-BHAR correlations for different AGN 
feedback models, for 
different 
epochs of the simulations. The best-fit parameters $\alpha$ and $\beta$ correspond 
to $\log{(BHAR)}=\alpha+\beta\times\log{(SFR)}$.\label{all} }
\begin{tabular}{@{}lcrrc}
\hline
Model & $\alpha$ & $\beta\;\;\;\;\;\;\;\;$ & $\chi^2$ & Epoch \\
\hline
SDH &     $-1.89\pm0.21$ & $-0.26\pm0.44$ & 0.05 & All \\
BS &      $-2.75\pm0.15$ & $0.20\pm0.28 $ & 0.83 & All \\
WT &      $-2.11\pm0.09$ & $0.47\pm0.15 $ & 9.67 & All \\
HPNK &    $-2.41\pm0.09$ & $0.09\pm0.20 $ & 4.90 & All \\
DQM &     $-1.66\pm0.06$ & $-0.25\pm0.16$ & 12.62 & All \\
ONB &     $-2.35\pm0.02$ & $-0.14\pm0.05$ & 27.06 & All \\
PNK0505 & $-2.13\pm0.06$ & $0.76\pm0.16 $ & 1.57 & All \\
SDH &     $-2.23\pm0.16$ & $-0.32\pm0.41$ & 10.99 & Pre-merger \\
BS &      $-2.93\pm0.29$ & $-0.13\pm0.41$ & 0.22 & Pre-merger \\
WT &      $-2.16\pm0.07$ & $0.28\pm0.22 $ & 6.75 & Pre-merger \\
HPNK &    $-2.42\pm0.10$ & $-0.63\pm0.28$ & 13.03 & Pre-merger \\
DQM &     $-2.44\pm0.31$ & $0.01\pm0.50 $ & 0.93 & Pre-merger \\
ONB &     $-2.54\pm0.04$ & $0.25\pm0.10 $ & 0.99 & Pre-merger \\
PNK0505 & $-1.98\pm0.06$ & $-0.01\pm0.22$ & 6.80 & Pre-merger \\
SDH &     $-2.03\pm0.28$ & $1.07\pm0.45 $ & 10.92 & Post-merger \\
BS &      $-2.46\pm0.20$ & $1.10\pm0.17 $ & 5.51 & Post-merger \\
WT &      $-1.35\pm0.26$ & $1.54\pm0.30 $ & 0.51 & Post-merger \\
HPNK &    $-1.94\pm0.23$ & $0.87\pm0.25 $ & 0.77 & Post-merger \\
DQM &     $-1.38\pm0.07$ & $0.61\pm0.24 $ & 3.49 & Post-merger \\
ONB &     $-2.25\pm0.00$ & $-0.01\pm0.02$ & 1.23 & Post-merger \\
PNK0505 & $-2.08\pm0.08$ & $1.25\pm0.19 $ & 3.65 & Post-merger \\

\hline
\end{tabular}
\end{table}

\subsection{Correlations of spatially decomposed SFRs and BHARs}

Given that the SFR and BHAR both have a strong dependence on the 
availability of cold gas, it is reasonable to expect stronger 
correlations between the nuclear SFR and BHAR than star formation in the 
extended regions of the galaxy. To define regions, we separate the nuclear 
and extended star formation by 1 kpc radial aperture cut-off. This is in 
agreement with the middle bin considered by \cite[hereafter 
DSR12]{DSR12} but slightly smaller in scale than the minimum 1.7 kpc 
value used by \cite{Lam13}, which varied up to 3.5 kpc with galaxy 
redshift, due to the fixed size of the SDSS spectroscopic fiber. We note 
that the correlations in the literature have been reported in a reversed 
form to \cite{Chen13}, i.e. as SFR$_{nuclear}$ $\propto$ 
(BHAR)${}^{\beta_1}$. For this format, DSR12 report an exponent of 
$\beta_1=0.61^{+0.15}_{-0.11}$ for the 1 kpc cut-off, while \cite{Lam13} 
report an exponent of $\beta_1=0.36\pm$0.04. Note that a decomposition 
of the DSR12 data into different aperture radii has a trend of 
decreasing exponent with increasing aperture radius, albeit at the cost 
of the smaller radii not including 24 $\mu$m continuum, which could lead 
to a systematic bias. For the extended star formation they find 
$\beta_1=0.57^{+0.28}_{-0.17}$, which is still somewhat higher than the 
nuclear correlation found by \cite{Lam13}. Taken together these two 
results do indicate a stronger correlation of nuclear star formation 
with the BHAR, than for extended star formation, as would be expected.

To calculate the results for the spatially decomposed regions we have 
followed the correlation approach of DSR12, but in the absence of errors 
on the SFR and BHAR measurements we apply a (log-space) least-squares 
approach to calculate the best-fit, rather than using the Bayesian 
methodology of DSR12. However, using this approach introduces some 
problems for the simulations in that exponentially low or zero values 
tend to influence trends strongly. While approaches such as ``Cook's D''
 can be used to determined which outliers weight most strongly, we have 
taken a conservative approach of removing zero values and any BHAR and 
SFR values lower than $10^{-6}$ \msol yr${}^{-1}$. While admittedly 
somewhat arbitrary, we believe this approach provides the best way of 
determining the correlations of low to moderate activity. The amount of 
data removed is given in Table \ref{nuclear}, and we denote whether data was removed 
due to low SFRs or BHARs by an $s$ or $b$ subscript. Only three models 
were impacted, WT, HPNK and PNK0505, mostly with less than 10\% of the 
data being impacted.

 To provide rough visual guidelines on the accuracy of the fits 
we have also calculated confidence bands.
Lastly, we note that the 
high activity episodes of the pre-merger galaxies are better analogues 
to the observed Seyfert sample (DSR12) than the post-merger 
remnant, but we include all data for completeness.

\begin{table}
\caption{The time averaged BHAR nuclear-SFR correlations for different 
AGN
feedback models, for
different
epochs of the simulations. Parameters are reversed compared to Table \ref{all}, in that
the best-fit parameters $\alpha_1$ and $\beta_1$ correspond
to $\log{(SFR)}=\alpha_1+\beta_1\times\log{(BHAR)}$.\label{nuclear} }
\tabcolsep=0.11cm
\begin{tabular}{@{}lrrcc}
\hline
Model & $\alpha_1\;\;\;\;\;\;\;\;$ & $\beta_1\;\;\;\;\;\;\;\;$ & \% outliers & Epoch \\
\hline
SDH &  $0.34\pm0.09$ & $0.19\pm0.03$ & 0 & All \\
BS &   $0.79\pm0.15$ & $0.32\pm0.04$ & 0 & All \\
WT &   $1.51\pm0.22$ & $0.86\pm0.08$ & $6_s$ & All \\
HPNK & $1.03\pm0.21$ & $0.59\pm0.07$ & $9_b$ & All \\
DQM &  $-0.12\pm0.08$ & $0.04\pm0.03$ & 0 & All \\
ONB &  $-0.98\pm0.28$ & $-0.37\pm0.11$ & 0 & All \\
PNK0505 & $0.50\pm0.08$ & $0.29\pm0.03$ & $10_b$ & All \\
SDH &  $0.56\pm0.10$ & $0.22\pm0.03$ & 0 & Pre-merger \\
BS &   $0.23\pm0.09$ & $0.08\pm0.02$ & 0 & Pre-merger \\
WT &   $0.47\pm0.09$ & $0.24\pm0.03$ & 0 & Pre-merger \\
HPNK & $0.32\pm0.10$ & $0.24\pm0.04$ & $9_b$ & Pre-merger \\
DQM &  $0.07\pm0.13$ & $0.08\pm0.05$ & 0 & Pre-merger \\
ONB &  $-0.91\pm0.36$ & $-0.34\pm0.14$ & 0 & Pre-merger \\
PNK0505 & $0.19\pm0.10$ & $0.13\pm0.04$ & $18_b$ & Pre-merger \\
SDH &  $0.12\pm0.17$ & $0.19\pm0.06$ & 0 & Post-merger \\
BS &   $0.84\pm0.21$ & $0.51\pm0.05$ & 0 & Post-merger \\
WT &   $2.22\pm0.18$ & $1.60\pm0.06$ & $17_s$ & Post-merger \\
HPNK & $1.21\pm0.37$ & $0.83\pm0.12$ & $8_b$ & Post-merger \\
DQM &  $1.39\pm0.12$ & $1.14\pm0.08$ & 0 & Post-merger \\
ONB &  $10.50\pm2.83$ & $4.75\pm1.26$ & 0 & Post-merger \\
PNK0505 & $1.26\pm0.10$ & $0.68\pm0.04$ & 0 & Post-merger \\

\hline
\end{tabular}
\end{table}


\subsubsection{Nuclear regions}\label{sec:nuc}

\begin{figure*}
\begin{center}
\includegraphics[width=2.\columnwidth]{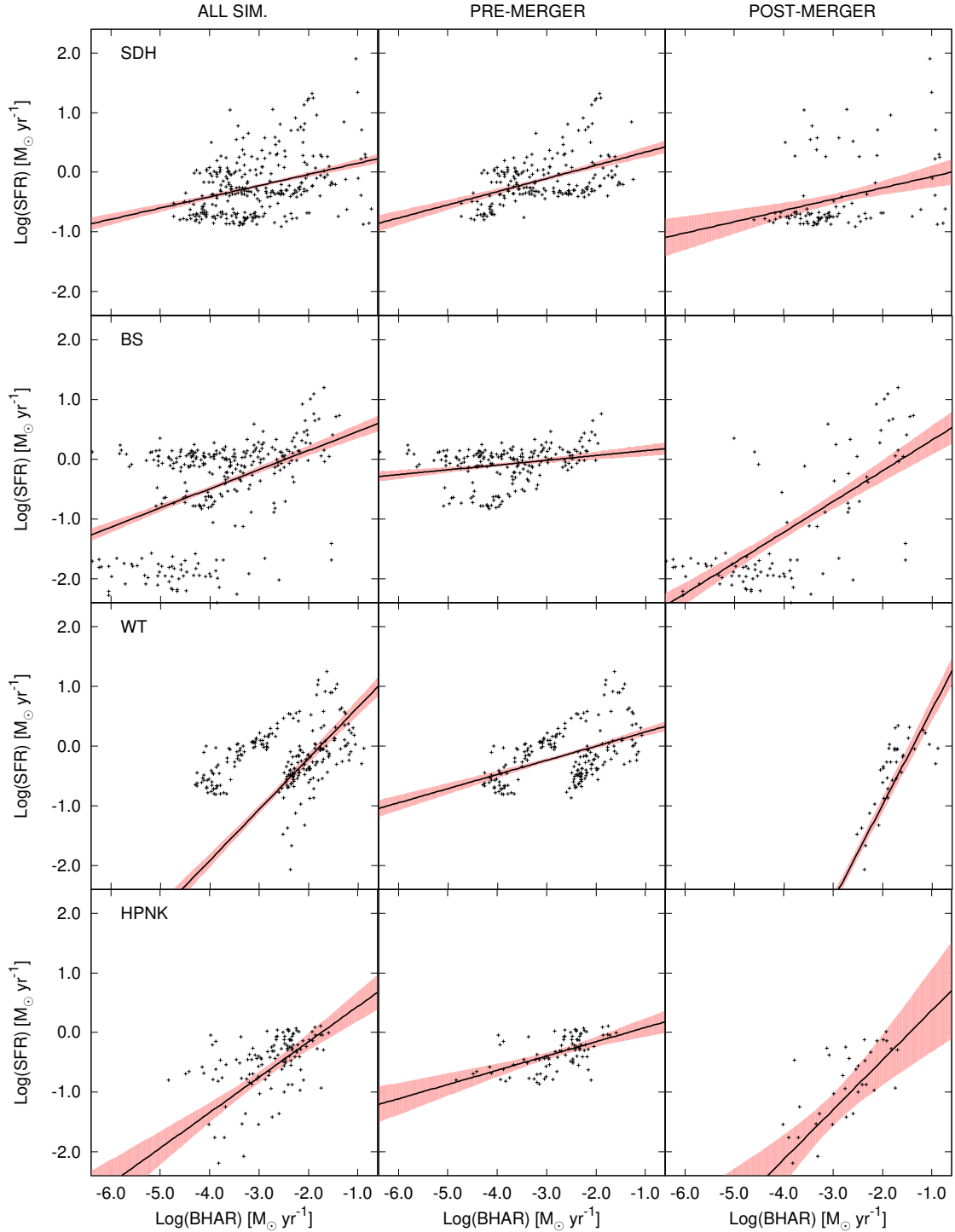}

\end{center}
\caption{Time averaged correlations between the BHAR and nuclear ($r<1$ kpc) SFR for the models that use Bondi-Hoyle
accretion or some variant of it (note the different orientation of axes compared to Figure \ref{bins1}). Each point
corresponds to the instantaneous SFR and BHAR values from outputs spaced 5 Myr apart within the simulations (10 Myr in the case of HPNK). Results are given for three different epochs, all the simulation,
the pre-merger evolution and the post-merger evolution. The filled red areas denote 95\% confidence bands around each fit.
\label{fig:nuc1}}
\end{figure*}

\begin{figure*}
\begin{center}
\includegraphics[width=2.\columnwidth]{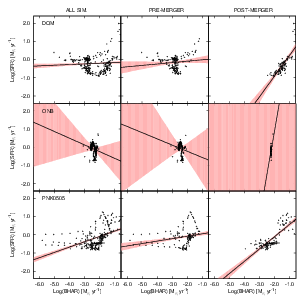}
\end{center}
\caption{Time averaged correlations between the BHAR and nuclear ($r<1$ kpc) SFR for the models that do not use Bondi-Hoyle
accretion or some variant of it. All other details are as for figure \ref{fig:nuc1}. Note that for the ONB model the narrow range
in BHAR means that the confidence bands are particularly large, the data would clearly be better fit if the axes were reversed.\label{fig:nuc2} 
 }
\end{figure*}

For this analysis, we present least-squares fits in Table \ref{nuclear}. The raw simulation data are plotted in
 \fig \ref{fig:nuc1} and 
\fig \ref{fig:nuc2}, along with the fits and confidence bands. 

With the exception of the ONB model, we find positive correlations 
between the SFR and BHAR for all models across all epochs. For the 
entire simulation period, we find an average for the positive slopes 
(i.e. excluding the ONB model) of $\beta_1=0.38^{+0.48}_{-0.34}$. This 
is somewhat lower than the value reported by DSR12 of 
$\beta_1=0.61^{+0.15}_{-0.11}$ but similar to the LaMassa \etal (2013) 
value of $\beta_1=0.36\pm0.04$. Of course individual models do however 
vary considerably away from this mean value with the WT model having the 
steepest slope of $\beta_1=0.86\pm0.08$ and the DQM model having the 
shallowest slope at $\beta_1=0.04\pm0.03$. 

For the post-merger systems, in agreement with the entire galaxy analysis,
we find correlations that are closer to linear
with an average and range across models of 
$\beta_1=0.83^{+0.78}_{-0.64}$. Notably, SDH is unusual in that its 
post-merger slope, $\beta_1=0.19\pm0.06$, is the same as the value for 
the entire simulation. This value is also the lowest slope for the 
post-merger systems. The high outlier is again WT with 
$\beta_1=1.60\pm0.06$. With exception of SDH and BS, the post-merger 
slopes are greater than DSR12.

For the positive slope pre-merger systems we find a shallow slope with 
comparatively little variation, with the mean and range across models 
being $\beta_1=0.17^{+0.07}_{-0.09}$. BS and DQM have the shallowest 
slopes with a $\beta_1=0.08$ value, while WT and HPNK share the steepest 
slope at $\beta_1=0.24$. For the pre-merger systems over half the 
evolution occurs with SFRs around or below 1\msol yr${}^{-1}$, while the 
BHAR evolves comparatively rapidly. The slight differences in the slope 
value can be traced to differences in SFR activity. For example, DQM 
stabilizes the galaxies against bar formation and thus keeps SFRs low, 
similarly BS has a small amount of early AGN activity that prevents 
higher star formation rates. These models show the shallowest slopes. 
However, models with episodes of somewhat higher SFRs, e.g. WT, SDH, 
HPNK where the bar mode has moderate strength, show slightly steeper 
slopes.

The post-merger slope observed for the WT model is also interesting in 
the context of the appearance of voids around the black hole (see 
\citealp{Bar13}), as it has the largest void of all the models. These 
voids, typically of size up to 1 kpc in radius, but which are ultimately 
dependent on resolution (see figure 13 in WT13b), can form as a product 
of the black hole influence/smoothing radius growing in size to 
encompass a sufficient number of neighbour particles. These voids can be 
produced in all models that follow the approach of increasing the black 
hole smoothing length to encompass a fixed number of neighbours (models 
SDH and BS, for example, both produce voids about 60\% the size of the 
WT model). In the merger simulations we describe, the void is formed in 
the WT model about 200 Myr from the final time. Thus the slope clearly 
has the potential to be impacted.

However, the formation of the void in the WT model is not merely the 
result of the black hole growing larger and larger. Following the merger, a 
large amount of gas is heated into a fountain-like process and falls 
back down on to the nuclear regions around the 1.2 to 1.3 Gyr point. 
Visual inspection suggests that the void is larger in this model partly 
because the infalling material has a notable amount of angular momentum 
and naturally settles at radii beyond 1 kpc.

To assess the impact of voids we have examined the SFR and BHAR 
data in the WT run. The very lowest SFRs in the WT model are below the 
$10^{-6}$ \msol yr${}^{-1}$ cut-off we employ and correspond to the 
epoch when the void has formed (approximately half the SFR values during 
this period are zero, interspersed with non-zero values between 
$10^{-5}$ \msol yr${}^{-1}$ to $10^{-3}$ \msol yr${}^{-1}$). Including 
the zero values in the least squares fit is not possible, but if we 
arbitrarily set the values to $10^{-10}$ \msol yr${}^{-1}$ this tilts 
the found power law slope to $\beta_1=2.38\pm0.21$, from 
$\beta_1=1.60\pm0.06$. This demonstrates, at least for the WT model,
that the void appears to have impacted the calculated correlation.

It is, however, important to determine if models without a significant 
void at all times can produce slopes that are equally steep, and whether 
there are any models with voids that, alternatively, produce shallow 
slopes. Firstly, the DQM model which has an extremely small void 
(essentially smaller than the simulation resolution), with 
$\beta_1=1.14\pm{0.08}$, the second steepest slope after the WT model. 
While the BS model, which has a void slightly smaller than WT, has 
$\beta_1=0.51\pm0.05$, a comparatively shallow slope.

We thus conclude that the impact of the voids on the computed 
correlations is not necessarily larger than other physical properties 
such as the feedback model, and the overall effect seems model dependent. 
Undoubtedly some element of control should be placed upon 
these voids to stop them becoming too large. Methods have been suggested 
elsewhere \citep{Bar13}.

\subsubsection{Outer regions}
\begin{figure*}
\begin{center}
\includegraphics[width=1.95\columnwidth]{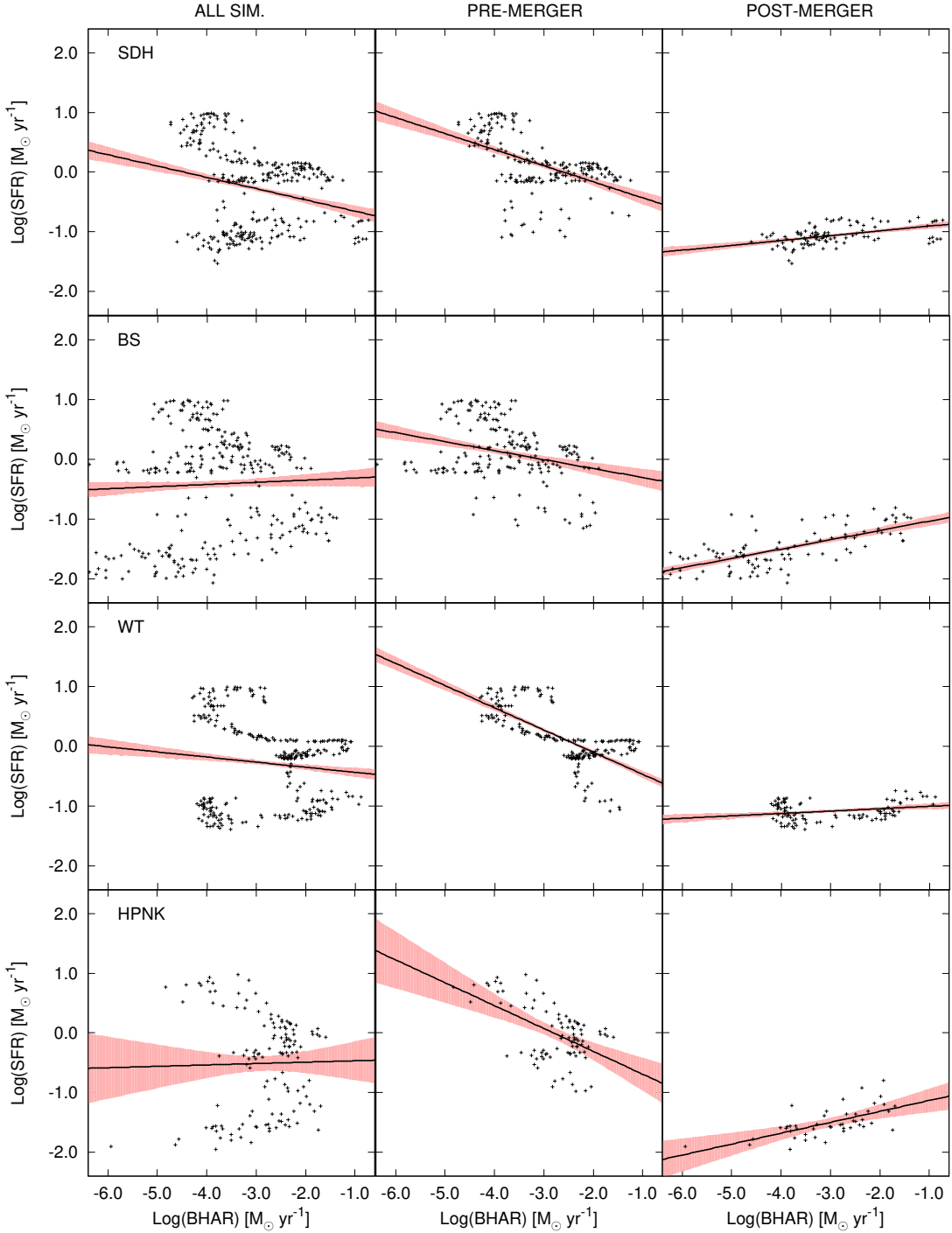}
\end{center}
\caption{Time averaged correlations between the BHAR and the ``outer" ($r>1$ kpc) SFR for the models that use Bondi-Hoyle
accretion or some variant of it (note the different orientation of axes compared to Figure \ref{bins1}). Results are given for three different epochs, all the simulation,
the pre-merger evolution and the post-merger evolution. The filled red areas denote 95\% confidence bands around each fit.\label{fig:ext1}}
\end{figure*}

\begin{figure*}
\begin{center}
\includegraphics[width=2.\columnwidth]{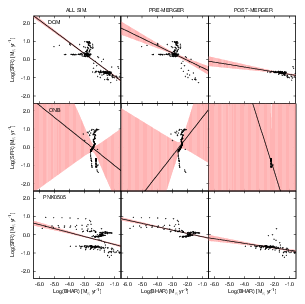}
\end{center}
\caption{Time averaged correlations between the BHAR and the ``outer" ($r>1$ kpc) SFR for the models that do not use Bondi-Hoyle
accretion or some variant of it. All other details are as for figure \ref{fig:nuc1}. Note that for the ONB model the narrow range
in BHAR means that the confidence bands are particularly large, the data would clearly be better fit if the axes were reversed. \label{fig:ext2}
 }
\end{figure*}

\begin{table}
\caption{The time averaged extended SFR-BHAR correlations for different AGN
feedback models, for
different
epochs of the simulations. Parameters are as in Table \ref{all}.\label{extended} }
\tabcolsep=0.11cm
\begin{tabular}{@{}lrrcc}
\hline
Model & $\alpha\;\;\;\;\;\;\;\;$ & $\beta\;\;\;\;\;\;\;\;$ & \% outliers & 
Epoch \\
\hline
SDH & $-0.84\pm0.13$ & $-0.19\pm0.04$ & 0 & All \\
BS & $-0.28\pm0.18$ & $0.04\pm0.04$ & 0 & All \\
WT & $-0.52\pm0.12$ & $-0.09\pm0.04$ & 0 & All \\
HPNK & $-0.44\pm0.28$ & $0.02\pm0.09$ & $9_b$ & All \\
DQM & $-1.52\pm0.06$ & $-0.62\pm0.03$ & 0 & All \\
ONB & $-1.67\pm0.57$ & $-0.66\pm0.24$ & 0 & All \\
PNK0505 & $-0.75\pm0.08$ & $-0.22\pm0.03$ & $10_b$ & All \\
SDH & $-0.70\pm0.11$ & $-0.27\pm0.03$ & 0 & Pre-merger \\
BS & $-0.46\pm0.14$ & $-0.15\pm0.03$ & 0 & Pre-merger \\
WT & $-0.85\pm0.08$ & $-0.37\pm0.03$ & 0 & Pre-merger \\
HPNK & $-1.07\pm0.18$ & $-0.38\pm0.06$ & $9_b$ & Pre-merger \\
DQM & $-0.84\pm0.14$ & $-0.40\pm0.05$ & 0 & Pre-merger \\
ONB & $2.71\pm0.59$ & $1.06\pm0.24$ & 0 & Pre-merger \\
PNK0505 & $-0.29\pm0.04$ & $-0.18\pm0.02$ & $18_b$ & Pre-merger \\
SDH & $-0.83\pm0.04$ & $0.08\pm0.01$ & 0 & Post-merger \\
BS & $-0.88\pm0.07$ & $0.16\pm0.02$ & 0 & Post-merger \\
WT & $-0.96\pm0.04$ & $0.04\pm0.01$ & 0 & Post-merger \\
HPNK & $-0.95\pm0.10$ & $0.18\pm0.03$ & $8_b$ & Post-merger \\
DQM & $-0.94\pm0.04$ & $-0.13\pm0.02$ & 0 & Post-merger \\
ONB & $-6.63\pm1.96$ & $-2.60\pm0.87$ & 0 & Post-merger \\
PNK0505 & $-0.98\pm0.05$ & $-0.13\pm0.02$ & 0 & Post-merger \\


\hline
\end{tabular}
\end{table}

Following the same analysis procedure as for the nuclear regions, we summarize the least-squares fits in Table \ref{extended}.
The raw simulation data are then plotted in
 \fig \ref{fig:ext1} and
\fig \ref{fig:ext2}, along with the fits and confidence bands.

It is immediately striking that the pre-merger slopes are almost all 
negative (ONB is again an exception, but is poorly fit due to the narrow 
range in BHAR), with a mean and range of 
$\beta_1=-0.29^{+0.14}_{-0.11}$. The most negative slope is the DQM 
model with $\beta_1=-0.40\pm0.05$, while the least negative (excluding 
ONB) is BS with $\beta_1=-0.15\pm{0.03}$. At this early stage of evolution
the extended SFR is comparatively unimpacted by events in the nuclear
region, and the Bondi-Hoyle variants have similar point distributions in
the SFR-BHAR parameter space. 

Across the entire simulation the mean and range of the slopes are given 
by $\beta_1=-0.18^{+0.22}_{-0.44}$. DQM has the most negative slope with 
$\beta_1=-0.62\pm{0.03}$ while the least negative slope is BS which is 
slightly positive at $\beta_1=0.04\pm0.04$. Comparatively few of the 
models agree within errors.

For the post-merger analysis, most simulations only have a very narrow 
range in SFR and, again excluding ONB, we find a mean and range of 
$\beta_1=0.03^{+0.15}_{-0.16}$. The Bondi-Hoyle variants (SDH, BS, WT 
and HPNK) have slightly positive slopes while DQM and 
PNK0505 have exactly matching negative slopes, $\beta_1=-0.13\pm0.02$. 
However, overall the variation in the post-merger slopes is 
comparatively small. ONB again remains an outlier because there is so 
little variation in the BHAR value.

The negative slope for the pre-merger systems (ONB excluded) primarily 
arises from events during the period of first periapsis. At this time 
the disc is still fairly gas rich while the black hole is has not grown 
significantly. This produces
 a relatively high SFR is accompanied by a very low BHAR, in turn 
placing points in the upper left of the parameter space that end up 
producing a negative correlation. DQM shows slightly different behaviour
at that time, but still produces a negative slope due to a number of
comparatively high BHAR events that are accompanied by low SFR values. 

The origin of the drop in the initial BHAR for models SDH, BS, WT, and 
HPNK can be traced to the change in the sound speed around the black 
hole. For these models there is an interaction between a high nuclear 
SFR and a small amount of thermal feedback from the AGN that drives up 
the sound speed of the gas surrounding the black hole. Because the ONB 
model injects feedback into the halo, while the DQM model uses a kinetic 
boost and has little initial feedback, neither of these models introduce 
a significant temperature change in the initial configuration. The DQM 
model does have a notable change in the density around the black hole, 
but because the influence radius changes in response, the ratio of the 
surface density to the angular velocity doesn't fall significantly, and 
the accretion rate remains comparatively constant at these early stages.

\subsection{Merger fractions and post-starburst black hole mass growth}\label{merge}

We next consider the contribution of mergers to the black hole mass and 
the growth of the black hole following the core merger at 980 Myr of 
evolution. It is worth emphasizing that the mass growth mode in the 
simulations is precisely determined by the assumed accretion law, be it 
Bondi-Hoyle or drag, or viscous accretion. 
Since the star formation algorithm 
does not account for mass loss into the ISM, which could subsequently be 
accreted on to the black hole, this mode of mass growth is not included.

For the merger fractions there is a maximum upper limit to the value
determined in the simulations. 
If two 
equal mass black holes merged and then there was no post-merger mass 
growth, the merger contribution is 50\%. Examining 
table \ref{massgrow} shows that models with little post-merger mass 
growth (e.g. WT, PNK0505) follow this trend and have 
comparatively high merger fractions. Models that 
exhibit extensive post-merger mass growth, particularly DQM and SDH, 
instead show comparatively small merger fractions. The 
range of values we find, namely from $\sim$10\% to $\sim$40\% are 
consistent with the growing expectation that for black hole masses below 
$10^9$ \msol$\,$ mergers do not play a dominant role in mass growth 
(e.g. \citealp{VC13,Du13,Ku13,Ang13}). However, what is perhaps 
surprising is the factor of four variation between models despite almost 
all of them matching the $M-\sigma$ relationship.

The post-merger mass growth rates do show considerable variation 
(approximately a factor of 3.5 between the lowest and highest values), 
and most are higher than the average post-starburst mass growth 
value of 5\% inferred by \cite{Wild10}. The merger simulation
is not however, markedly distinct from their chosen sample.
 The simulated black holes masses are at the upper limit of their 
inferred mass range of $10^{7.5}$\msol~ and the merger remnant stellar 
morphology is a very flattened ellipsoid that nonetheless does fall 
above their stellar mass surface density cut-off of 
$\mu^*>3\times10^8$\msol kpc${}^{-2}$. Black hole luminosities for the 
merger also peak within, but at the upper end, of their range at 
$10^{44}$ erg s${}^{-1}$.

\begin{table}

\caption{Final black hole masses, post-merger mass increase ratios and 
merger mass fractions for the different models. The ONB model is omitted as
for reasons noted at the beginning of section \ref{results}. \label{massgrow}}
\begin{tabular}{@{}lccc}
\hline
Model & Final mass/\msol & Mass Ratio & Merger fraction\\
\hline
BS & $5.56\times10^6$ & 2.55 & $0.16^\dagger$\\
SDH & $2.52\times10^7$ & 3.80 & 0.13 \\
WT & $2.75\times10^7$& 1.36 & 0.40 \\
HPNK & $8.80\times10^6$ & 1.20 & 0.42 \\
DQM & $3.32\times10^7$ & 4.94 & 0.08 \\
PNK0505 & $3.81\times10^7$ & 1.66 & 0.36 \\
\hline
\end{tabular}\\
{${}^{\dagger}$Note the merger fraction is estimated for the BS model
as the black holes did not actually merge.}\\
\end{table}

\section{Conclusion}
\label{conclusion}

We have presented a detailed analysis of the evolution of AGN feedback 
models in the BHAR-SFR parameter space and contrasted the time-averaged 
trends in this space to observed relationships for ensembles of 
galaxies. In addition to models considered 
in WT13a and WT13b we have also added an additional model, described in HPNK12. 
Our principle conclusions are:
\begin{itemize}

\item For the parameters we considered, the revised accretion model of 
HPNK produces significant early growth in the black hole masses, but 
produces considerably less growth at late times. The resulting final 
black hole mass is 1.4 standard deviations below the mean of 
the \cite{Gul09} M-$\sigma$ relationship, but still higher than other notable
models e.g. BS. Due to the lack of late time growth in the 
black hole mass, this model also has the largest mass contribution from 
mergers, albeit only slightly larger than the WT model.

\item Evolution of a single merger system in the SFR-BHAR parameter 
space is highly complex even when averaged over 20 Myr periods. While a 
number of the models, especially those using variants of Bondi-Hoyle 
accretion, do follow qualitatively similar evolution, namely a vertical 
rise followed by a diagonal decay to lower SFR and BHAR values, the 
precise quantitative behaviours can be distinctly different. Notably,
none of the models reproduces the inferred evolution from \cite{Wild10}, 
but without an ensemble of merger simulations of varying mass this 
result shouldn't be over-interpreted.

\item When converted into time-averaged correlations, the SFR-BHAR 
evolution manifests in expected ways. The pre-merger and full simulation 
correlations are generally flat, but the post-merger evolution for all 
models bar ONB, shows a distinct positive correlation, with some models 
being close to linear. However, the normalization of these periods of 
evolution lies above both the \cite{Chen13} and SFR/500 relationships.

\item Breaking the star formation into nuclear and extended components
reproduces the qualitative behaviour observed in observational work, namely that there
is a stronger correlation between the nuclear SFR and BHAR, than there is between
the outer SFR and the BHAR. There is also a distinctly stronger correlation for
the post-merger nuclear SFR than the pre-merger, which would be interesting to
probe observationally at the ensemble level. The different models do not, however,
favour one particular observational result, although the mean of the models over
the entire simulation, $\beta_1=0.38^{+0.48}_{-0.34}$, is surprisingly close to the LaMassa \etal (2013) value of $\beta_1=0.36\pm0.04$.

\item The models show significant variation in the contribution of mergers
to the final black hole mass. This variation is directly attributable to the amount
of mass growth that occurs post-merger: those models with little mass growth (e.g. WT, PNK0505) obviously
have large merger fractions. The post-merger mass growth values are usually considerably
larger than that derived by \cite{Wild10}, although the simulated system is
very much at the upper end of the mass range they consider. There may also be subtle timing issues here
related to when the starburst occurs relative to the main merger.  

\end{itemize}

While resolution issues are significant in AGN models, it is also equally important to 
understand the variance in evolution in the SFR-BHAR parameter space that occurs as a
result of different merger trees. This could be examined effectively through multiple
zoom simulations, although new high resolution uniform volume simulations have reached
the point where similar mass resolution to that considered here can be achieved. Adaptive-mesh refinement techniques 
also provide
an interesting alternative route to high resolution, although mass
resolution in the collisional part of the simulation must be carefully
considered against the extremely high spatial resolution that can be achieved.
 
Adding a full merger tree overcomes a number of issues (e.g. \citealp{JM13}) related
to the dynamics of isolated pairs of galaxies especially for lower mass systems. One possible
area in which there may well be a distinct impact is related to the overall time AGN
spend at a given luminosity (which we shall call the ``activity function''), which can be used in conjunction with halo population functions to estimate
the luminosity function of AGN. Even though we have included hot halos in our models, there is
no representation of accretion of lower mass halos that would typically be present in cosmological
environments. Undertaking a comparison of the activity function variance in different cosmological environments is clearly
an important next investigation because determining whether any of the models spend an appreciable amount of time
at lower luminosities, to match implied observational results, is not yet well understood. In the theme of the current
paper it is also worth investigating whether any of the models has a notable tail 
down to low luminosity in the isolated simulations. The model that stands out in this regard is PNK because of the 
presence of the exponential decay.

Lastly, while the current study has examined numerous different approaches to
AGN feedback, it is important to remember that the evolution in the
SFR-BHAR parameter space is also influenced by the precise SFR and stellar
feedback algorithm. To this end, examining the impact of effective equation
of state approaches on the evolution is also an important next step and one we plan to 
address with higher resolution simulations.


\section*{Acknowledgements}
We thank the anonymous referee for comments that helped improved the clarity and focus
of the paper.
RJT sincerely thanks Professors Carlos Frenk and Richard Bower, plus the members of the 
Institute of 
Computational Cosmology at the University of Durham for 
hosting him during the early stages of this work. JW was supported by NSERC and Saint 
Mary's University while the simulations used in this work were run, while CM was supported
by an NSERC summer studentship. RJT is supported by a Discovery Grant 
from NSERC, the Canada Foundation for Innovation, the Nova Scotia 
Research and Innovation Trust and the Canada Research Chairs programme.
Simulations were run on the CFI-NSRIT funded {\em St. Mary's 
Computational Astrophysics Laboratory}. 

\bibliography{mybib}

\label{lastpage}
\end{document}